\documentclass[aps,prd,amsmath,reprint]{revtex4-1}
\usepackage{graphicx}
\newcommand{\abs}[1]{\left|{#1}\right|}
\newcommand{\ket}[1]{\left|{#1}\right>}
\newcommand{\bra}[1]{\left<{#1}\right|}
\allowdisplaybreaks[1]
\begin{document}
\title{A quantum field theoretical detector model\\for probing the Unruh effect}

\author{Franz Thoma}%
	\email{f.thoma@physik.uni-muenchen.de}
	\affiliation{Arnold Sommerfeld Center for Theoretical Physics,\\ Ludwig-Maximilians-Universit\"at,\\
			Theresienstra{\ss}e 37, 80333 Munich, Germany}


\begin{abstract}
We consider a particle detector model on 1+1-dimensional Minkowski space-time that is accelerated by a constant external
acceleration $a$. The detector is coupled to a massless scalar test field.
Due to the Unruh effect, this detector becomes excited even in Minkowski vacuum with a probability
proportional to a thermal Bose-Einstein distribution at temperature $T=\frac{a}{2\pi}$ in the detector's energy gap.
This excitation is usually said to happen upon detection of a Rindler quantum, which is defined by having positive
frequency with respect to the detector's proper time instead of inertial time.
Using Unruh's fully relativistic detector model, we show that the process involved is in fact spontaneous excitation
of the detector due to the interaction with the accelerating background field $E=ma$.
We explicitly solve the Klein-Gordon equation in the presence of a constant background field $E$
and use this result to calculate the transition probability for the detector to become excited on Minkowski vacuum.
This transition probability agrees with the Unruh effect, but is obtained without reverting to the concept of Rindler quantization.
\end{abstract}

\maketitle

\section{Introduction}
Ever since Fulling~\cite{Fulling73}, Davies~\cite{Davies74} and Unruh~\cite{Unruh76}
discovered that the Minkowski vacuum state of a scalar quantum field
is a thermal state with respect to the Rindler vacuum and therefore contains particles, there has been an ongoing discussion
on whether or not and how these particles can be observed.
The usual derivation of the Unruh effect
(using Bogolyubov transformations to connect the inequivalent field quantizations in Minkowski and Rindler coordinates,
respectively) does not feature the notion of a detector.
Instead, the particles are associated with an (accelerated) observer, where there is no exact (physical) definition
both of the term \emph{observer} and how to define \emph{associated}. It is just supposed that any observer,
be it a particle detector or some macroscopic system like a human being, will perceive Minkowski vacuum as a thermal state
if its proper time is proportional to that of an accelerated (rather than inertial) particle.

In order to give a more practical definition to these terms, Unruh (1976~\cite{Unruh76}) and DeWitt (1979~\cite{DeWitt79})
proposed to use a simplified model particle detector, commonly referred to as the ``Unruh'' or ``Unruh-DeWitt'' detector model,
consisting of a two-level Schr\"odinger system enclosed in an arbitrarily small box
which is said to have detected a particle if it has made a transition to its excited state.
This detector is set up to detect Rindler particles by switching from inertial (Minkowski) time to
Rindler time, the proper time of an accelerated particle.
Then the probability for the detector to switch to its excited state is found to be proportional
to a Bose-Einstein factor at temperature $T=\frac{a}{2\pi}$ in agreement with the prediction from te Unruh effect.
However, when analyzing the final state of the entire system (rather than only of the detector),
it was found that along with the excitation of the detector, a Minkowski particle is emitted~\cite{Unruh76,UnruhWald84}.

This assertion was attacked by Grove, who computed the expectation value of the stress-energy energy tensor
at late times and claimed that the increase in the stress-energy of the test field is merely an artifact
of the reduction of the full ``out'' state to a product state~\cite{Grove86}.
He concluded that the change in the energy of the test field does not correspond to a dynamical process;
consequently, for a macroscopic detector (i.e. an ensemble of model detectors) the effect should not be present
since reduction effects should be ``averaged out'' over the particle number.

Audretsch~and~M\"uller~\cite{Audretsch94b} resolved this issue by computing the full out state of the
coupled system (detector and test field) up to second order in the interaction.
They analyzed both the Minkowski energy and the Minkowski particle content of the test field
in the three cases that (a)~no measurement of the detector state was made,
that (b)~the detector was found in the excited state and
that (c)~the detector was found in the ground state.
The result is that while the Minkowski energy density only changes due to a reduction effect
and will stay zero in the left Rindler wedge when averaged over a macroscopic number of detectors,
the Minkowski particle number will indeed increase even if no measurement (state reduction) is made on the detector.

What still remains is the question whether or not Rindler particles can be considered ``real particles''.
On one hand, one may adopt the point of view that the concept of a ``particle'' itself depends on the observer,
or rather the reference frame one is working in. This is the view held by Wald~%
\cite{Wald94}. 
The Unruh-DeWitt detector model is also based on this view:
The choice of coordinates in which the the Green's functions are computed determines which particles,
Rindler or Minkowski, the detector is sensitive to, working with both particle definitions on equal footing.

On the other hand, however, DeWitt pointed out that in flat space-time Minkowski vacuum is distinct from all other vacua
in being the only vacuum state with the expectation value of the energy-momentum tensor being identically zero,
which is required in flat space time in order to be consistent with general relativity~\cite{DeWitt79}.

We want to emphasize that this is a problem peculiar to the Unruh effect, not to inequivalent field quantizations per se.
When compared for example to the Hawking~effect~\cite{Hawking74,*Hawking75}, which relies on the same Bogolyubov formalism,
we observe a huge conceptional difference: 
For the Hawking effect, we use different field quantizations in different regions of space-time
(past and future null infinity). Therefore, the notion of what is a particle changes over time,
but at each point in space-time there is a meaningful and unambiguous definition for the particles that contribute
to the stress-energy tensor.
For the Unruh effect on the other hand, we use two inequivalent field quantizations
at the same point in space-time, so that the concept of particles only becomes meaningful with the specification
of an associated observer, and it remains a priori unclear which quanta conribute to the stress energy tensor.

In this paper we are going to resolve this issue using the other detector model proposed by Unruh,
a fully relativistic detector~\cite{Unruh76}.
This model has gained rather little attention so far, mostly due to the fact that Unruh himself claimed
this model would ``display the same phenomena''~\cite{Unruh76} as the simpler model (which we shall prove correct).
However, being a quantum field theoretical model, it is better suited for probing relativistic field theoretical phenomena
such as the Unruh effect:
Rather than just switching coordinates, this model explicitly incorporates a constant background field to
account for the external acceleration.

We are going to use this relativistic detector model to show that the process of Rindler particle detection is
in fact just spontaneous excitation of the detector due to the interaction with the accelerating background field.
This interpretation is justified since we do not need to revert to the concept of Rindler quanta.
Another benefit of this method is that we can drop the \emph{assumption}~%
\cite{[It was pointed out by ][{ that ``this is a physical hypothesis, and should not simply be taken for granted''}]Sciama81}
that the detector measures frequencies
with respect to its proper time: By incorporating the ``motor'' of the external acceleration explicitly into
the detector model, we show that this is just a consequence of energy and momentum conservation.

\section{The accelerated quantum field\label{sec:AcceleratedField}}
In this section we are going to discuss the behavior of a constantly accelerated massive scalar quantum field~%
\cite{[The framework discussed here was proposed by ]Brout95b,*[See also ][ for a more in-depth view on this subject]Brout95a}
of mass $m$. The acceleration $a$ shall be generated by an external electric field $E=ma$.
We will encounter that the standard solutions incorporate non-perturbative pair production similar to the Schwinger effect
in quantum electrodynamics.
The mode expansions in terms of parabolic cylinder functions shall be useful later for computing transition amplitudes
for the accelerated detector.

We start with the classical Hamilton function of a massive relativistic particle in a potential $V(x)=-Ex$,
\[H=\sqrt{p^2+m^2}-Ex.\]
The linear potential results in a constant force $E=ma$ with $m$ being the mass of the particle and $a$ the acceleration.
We can think of $E$ as an external electric field with unit charge $e=1$.
The orbits of this Hamiltonian system are the well-known hyperbolic trajectories of constantly accelerated particles,
\begin{align*}
x(t)=\frac{1}{a}\sqrt{1+a^2(t-t_0)^2}+x_0.
\end{align*}
The turning point is determined by the conserved total energy $\mathcal{E}$ of the system,
$x(t=t_0)=1/a+x_0=(m-\mathcal{E})/E$.

The usual quantization procedure yields the modified Klein-Gordon equation for a complex scalar field $\varphi$:
\begin{equation}\label{eq:KGEq}
0=(i\partial_t+Ex)\varphi+\partial_x^2\varphi-m^2\varphi =: (\mathcal{D}_\nu\mathcal{D}^\nu-m^2)\varphi
\end{equation}
with the covariant derivatives $\mathcal{D}_t=\partial_t-iEx$, $\mathcal{D}_x=\partial_x$.
This is the equation of motion for a charged field (unit charge $e=+1$) in a constant external electric field $E$.
The corresponding action reads
\begin{equation}\label{eq:KGAction}
S=\frac{1}{2}\int d^2x\ \bigl[\mathcal{D}_\nu^*\varphi^* \mathcal{D}^\nu\varphi + m^2\varphi^*\varphi \bigr],
\end{equation}
from which the field equation \eqref{eq:KGEq} and the corresponding anti-particle equation follow
via variation with respect to $\varphi$ and $\varphi^*$, respectively.

Since the external field is constant, in particular stationary, we can expand the field in terms of stationary modes
\[\varphi_\omega(x,t) = e^{-i\omega t} \varphi_\omega(x)\]
in order to reduce equation \eqref{eq:KGEq} to a stationary equation
\[\partial_x^2\varphi_\omega = m^2\varphi_\omega - (\omega+Ex)^2\varphi_\omega.\]
This equation is solved by the parabolic cylinder function~\cite{AS65} 
\begin{align*}
E_\mu\bigl(\sqrt{2E}(x+\omega/E)\bigr),&& E_\mu^*\bigl(\sqrt{2E}(x+\omega/E)\bigr)
\end{align*}
with the dimensionless quantity $\mu=\frac{m^2}{2E}$ being a measure for the strength of the external field
compared to the mass. The total energy of the system $\omega$ enters in the argument as translation term
and determines the classical turning point $x_0=\frac{m-\omega}{E}$.
This means in particular that $\omega$ is not bounded from below and may take also negative real values.
This is possible due to the fact that the potential energy compensates the lower total energy
so that for $x\geq x_0$ the kinetic energy is still positive
(at the turning point $x_0$, the kinetic energy takes its minimum $\omega_\text{kin} = \omega-Ex_0 = m$;
beyond that point, for $x<x_0$, particles are classically forbidden). 

\begin{figure}[t]
\includegraphics[height=1.0\columnwidth,angle=-90]{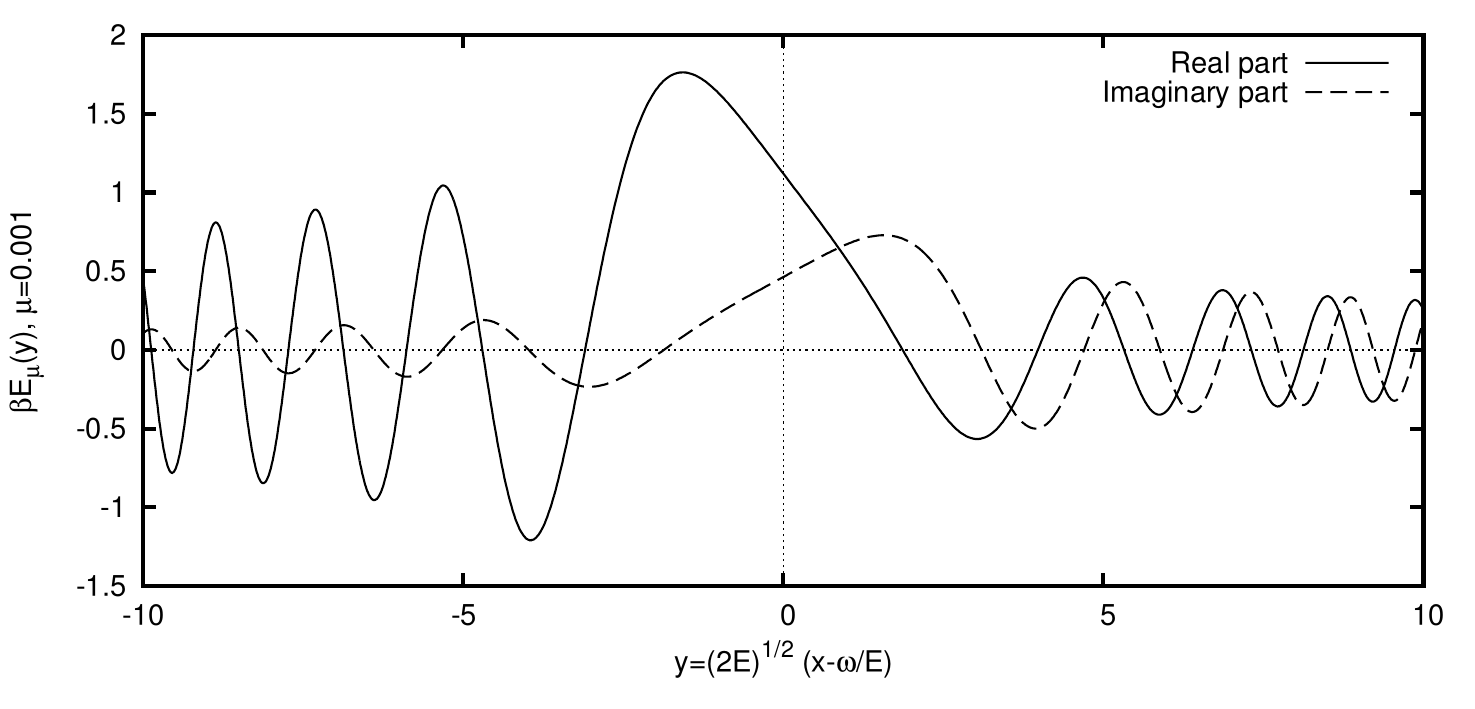}
\caption{\label{fig:ParabolicCylinder}%
A plot of the parabolic cylinder function $E_\mu(y)$ for $\mu=0.001$.
}
\end{figure}

Fig.~\ref{fig:ParabolicCylinder}%
\begin{figure*}
\includegraphics[height=1.0\textwidth,angle=-90]{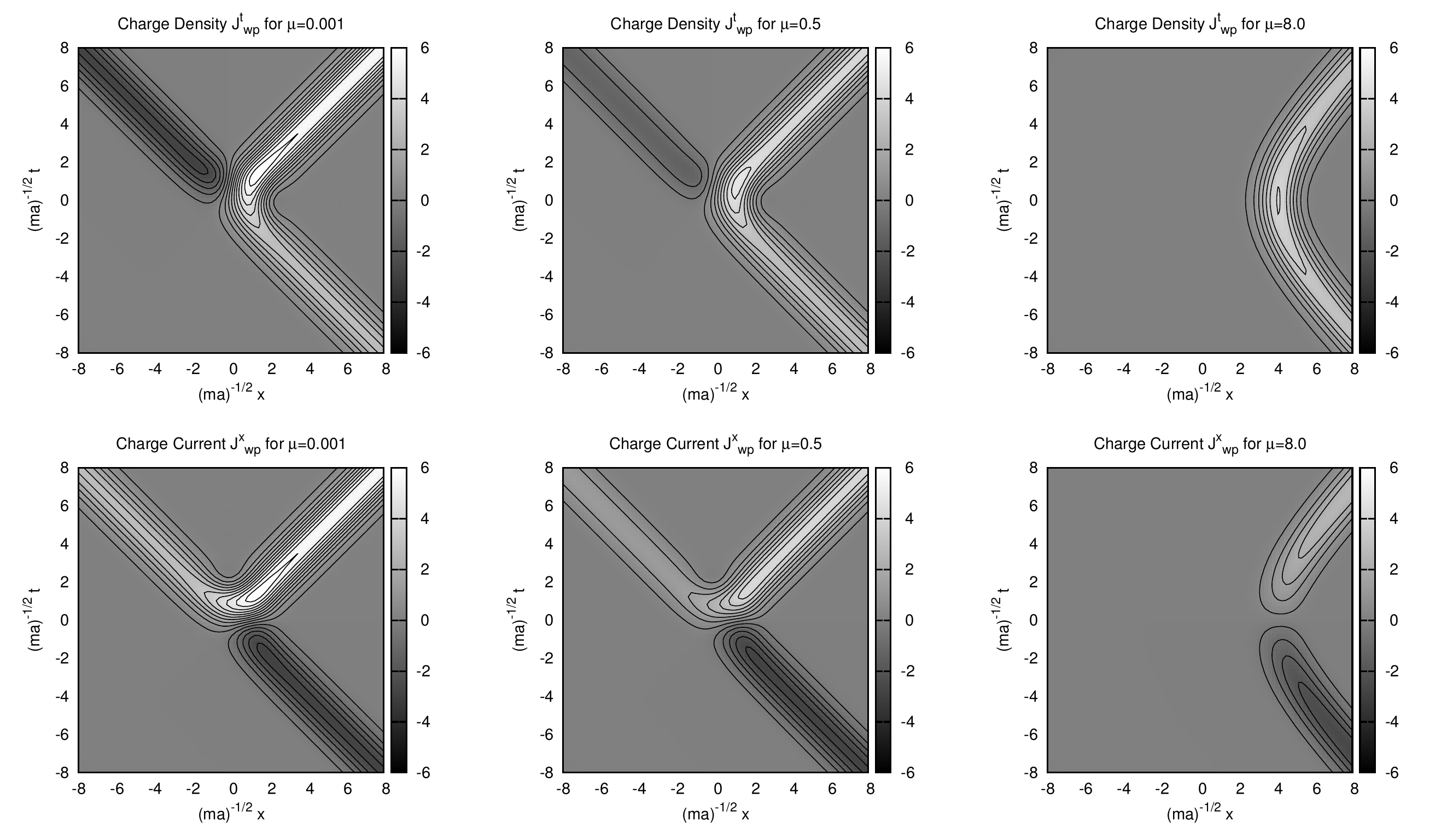}
\caption{\label{fig:ChargeDensities}%
	Charge density and charge current of the wave packet solution
	$\varphi_\text{wp}=\int d\omega\ e^{-\frac{\omega^2}{2E}} E_\mu^*\bigl(-\sqrt{2E}(x+\omega/E)\bigr)$
	for $\mu=0.001,0.5,8.0$.
	The contours are drawn for $J/E=\ldots,-0.9,-0.3,0.3,0.9,\dots$
	In the upper row (charge density), white corresponds to positive charges (particles) and
	black to negative charges (antiparticles).
	In the lower row (charge current), white corresponds to right-moving particles or left-moving antiparticles and
	black to left-moving particles.\\
	We observe an incoming particle that is reflected back to $+\infty$ (the prevalent process at low acceleration, here $\mu=8$)
	and the creation of a pair (anti-particle goes to the left, particle to the right),
	which can be seen at higher accelerations (for $\mu=8$, the charge density of the created anti-particle
	is more than 22 orders of magnitude smaller than for the incoming particle:
	$\abs{\beta}^2=e^{-2\pi\cdot8.0}=1.48\cdot10^{-22}$, too small to be seen in this plot).
}
\end{figure*}
shows a plot of the solution $E_\mu(y)$ for $\mu=0.001$ where we use the dimensionless
variable \[y=\sqrt{2E}(x+\omega/E)\] to make the discussion independent from the energy $\omega$.
For $y\to\infty$ the solution describes a single right-moving particle.
The relation
\begin{equation}\label{eq:LRRelation}
E_\mu(-y)=i\sqrt{1+e^{2\pi\mu}}E_\mu^*(y)-ie^{\pi\mu}E_\mu(y)
\end{equation}
shows that at $y\to-\infty$ there is a superposition of two modes travelling in opposite directions.

To sort these terms into particles and antiparticles, as well as ingoing and outgoing modes, we first observe
that all the modes at $y\rightarrow-\infty$ have to be antiparticles and all modes at $y\rightarrow\infty$ particles:
The classical turning points are at $y=\sqrt{\frac{2m}{a}}=2\sqrt{\mu}$ for particles
and $y=-2\sqrt{\mu}$ for antiparticles. The region to the left is classically forbidden for
particles, the region to the right for anti-particles. In these regions, the parts of the mode function corresponding to
(anti-)particles do not oscillate, they decay exponentially
(the middle region, $-2\sqrt{\mu} \le y \le 2\sqrt{\mu}$, is classically forbidden for both particles and antiparticles).

Looking at the charge flux, we see that for $E_\mu(y)$ we have positive flux, i.e.\ we interpret the
$y\rightarrow\infty$ part as a particle flux to the right, or outgoing particle flux.
Of course, the charge flux of antiparticles has opposite sign, so that positive charge flux corresponds to
right-moving particles and left-moving antiparticles.
Since both $y\rightarrow-y$ and $E_\mu\rightarrow E_\mu^*$ will change the direction of the flux,
this means for the interpretation of $E_\mu(y)$:
The $y>0$/right sector corresponds to an outgoing particle, $y<0$/right to an outgoing antiparticle,
$y<0$/left to an ingoing antiparticle.

We rearrange equation \eqref{eq:LRRelation} in terms of past and future modes,
\begin{equation}\label{eq:PFRelation}
\begin{aligned}
E_\mu^*(-y)&=ie^{-\pi\mu}E_\mu^*(y) &&+ \sqrt{1+e^{-2\pi\mu}}E_\mu(-y)\\
&=\beta E_\mu^*(-y) &&+ \alpha E_\mu(y).
\end{aligned}
\end{equation}
The left hand side describes a single particle in asymptotic past, an ingoing particle.
For asymptotic future, the right hand side shows that there are $\abs{\alpha}^2$ particles and
$\abs{\beta}^2=1-\abs{\alpha}^2$ anti-particles in the out-sector: the original particle plus a particle-antiparticle pair.
The strength of the pair production $\abs{\beta}^2=e^{-2\pi\mu}$ depends on the strength of the electric field:
For a weak electric field ($E\ll m^2$ or $\mu\gg1$), pair production is exponentially suppressed;
for a strong field on the other hand, $\abs{\beta}^2$ approaches one as $\mu\to0$.
Plotting charge density and charge current for a wave packet $e^{-\frac{\omega^2}{2E}}$ centered around $\omega=0$
(fig.~\ref{fig:ChargeDensities})
gives an accurate illustration of the pair creation process.

The correct normalization is obtained from the charge current density
\[j_\nu=i\bigl(\varphi\mathcal{D}^*_\nu\varphi^*-\varphi^*\mathcal{D}_\nu\varphi\bigr)\]
which is conserved by the internal $U(1)$ symmetry of the action \eqref{eq:KGAction}, $\partial_\nu j^\nu=0$.
For our stationary solution $\varphi_\omega(x)=\frac{1}{N}E_\mu^*(y)$, the current density $j_x$ is constant,
\[j_x=-2\frac{\sqrt{2E}}{N^2}.\]
This current arises from the opposite charges of the created pair that travelling in opposite directions
(the incident particle does not contribute since it is reflected back to $+\infty$ so that the net current is zero).
So $j_x$ needs to be proportional to the number of pairs created:
\[j_x\overset{!}{=}-\abs{\beta}^2\]
(assuming unit current density for a single particle) and therefore $\frac{1}{N}=\frac{\beta}{(8E)^{1/4}}$.
The standard ``in-particle'' mode is therefore
\[\varphi_\omega(x,t)=(8E)^{-1/4}\,\beta\,e^{-i\omega t}\,E_\mu\bigl(-\sqrt{2E}(x+\omega/E)\bigr)\]

The second quantized field operator can now be readily expanded in terms of past and future modes:
In asymptotic past, we have $\varphi_\omega(x,t)$ for ingoing particles and $\varphi_{-\omega}^*(-x,t)$
for ingoing anti-particles:
\begin{subequations}\label{eq:FieldExpansion}
\begin{equation}
\varphi(x,t)=\int_{-\infty}^\infty d\omega\ \Bigl[ a^\text{in}_\omega\, \varphi_\omega(x,t) 
			+ {{}\hat{a}^\text{in}_\omega}^\dagger\, \varphi^*_{-\omega}(-x,t)\Bigr]
\end{equation}
In asymptotic future, we expand in terms of outgoing modes $\varphi_\omega^*(x,t)$ and $\varphi_{-\omega}(-x,t)$:
\begin{equation}
\varphi(x,t)=\int_{-\infty}^\infty\! d\omega\, \Bigl[ a^\text{out}_\omega\, \varphi^*_\omega(x,t)
			+ {{}\hat{a}^\text{out}_\omega}^\dagger\, \varphi_{-\omega}(-x,t)\Bigr]
\end{equation}
\end{subequations}
The creation and annihilation operators for particles ($a^\dagger,a$) and antiparticles ($\hat a^\dagger, \hat a$)
obey the usual commutation relations.

\section{The relativistic Unruh detector}
The relativistic detector model introduced by Unruh~\cite{Unruh76} consists of three relativistic quantum fields:
The detector fields $\Phi$ and $\Psi$ corresponding to a detector particle in its ground and excited state, respectively,
and a massless scalar test field $\psi$. The action of the detector model
\begin{equation}
\begin{aligned}
S=\frac{1}{2}\int d^2x\ \Bigl[ &\mathcal{D}^*_\nu\Phi^\dagger \mathcal{D}^\nu\Phi &&+ M^2 \Phi^\dagger\Phi \\
		\Big. +{} &\mathcal{D}^*_\nu\Psi^\dagger \mathcal{D}^\nu \Psi &&+ \bar{M}^2 \Psi^\dagger\Psi \\
		\Big. +{} &\partial_\nu\psi\partial^\nu\psi \\
				&&&{}+ \epsilon \bigl( \Phi^\dagger\Psi+\Psi^\dagger\Phi \bigr)\psi\Bigr]
\end{aligned}
\end{equation}
contains the kinetic and mass terms of the three fields as well as the minimal interaction term.
We use the covariant derivatives from the action \eqref{eq:KGAction} to couple the detector fields to an external
background field $E$; consequently, $\Phi$ and $\Psi$ have to be complex fields.
The mass $\bar M$ of the excited detector should be at least slightly bigger than $M$; this mass gap induces an energy gap
between the $\Phi$ and $\Psi$ states which determines the energy the detector is senitive to.
The test field $\psi$ contains the particles we want to detect and is not coupled to the background field.
For simplicity, we assume a massless real field.

The detector model works as follows: A detector, represented by a $\Phi$ particle in asymptotic past,
is said to have ``detected'' a particle if it passes over to its excited state,
i.e. a $\Psi$ particle is found in asymptotic future instead.
In first order perturbation theory, the form of the interaction allows only two processes corresponding to ``detection'':
$\Phi\psi\to\Psi$ (test particle absorption) and $\Phi\to\Psi\psi$ (spontaneous excitation/test particle emission).
Energy and momentum conservation restricts the energy $\omega$ of the $\psi$ particle:
We find that the mass gap of the detector
\begin{equation}\label{eq:MassGap}
\omega_\text{d}=\frac{\bar M^2-M^2}{2M}
\end{equation}
determines the energy $\omega$ of the $\psi$ particle in the rest frame of the detector,
\begin{equation}\label{eq:InertialEnergy}
\omega=\pm\omega_\text{d}
\end{equation}
(in any other reference frame, $\omega_\text{d}$ is multiplied by a redshift factor).

In the absence of the background field ($E=0$ or $\mathcal{D_\nu}=\partial_\nu$), energy and momentum conservation
only allow the former transition ($\Phi\psi\to\Psi$ with $\omega=+\omega_\text{d}$);
the latter one, spontaneous emission of a test particle, is forbidden since $\omega=-\omega_\text{d}$ would have to be negative.

For $E\neq0$, however, $\omega=+\omega_\text{d}$ is also possible in the case of spontaneous excitation:
The energy required to create a $\psi$ particle is supplied by the external field.
This was already predicted by Unruh~\cite{Unruh76}, but the theory of the accelerated quantum field
from section \ref{sec:AcceleratedField} allows us to actually compute the transition amplitude.

\subsection{Spontaneous excitation\label{sec:SpontExcitation}}
We use the field expansion \eqref{eq:FieldExpansion} for the accelerated fields $\Phi$ and $\Psi$ and the usual Fourier mode
epansion for the test field $\psi$:
\begin{align*}
\Phi(x,t) &= 
	\begin{cases}
	\displaystyle\int_{-\infty}^\infty\!\! d\Omega\, \Bigl[ a^\text{in}_\Omega \varphi_\Omega(x,t)
				+ {{}\hat{a}^\text{in}_\Omega}^\dagger \varphi^*_{-\Omega}(-x,t) \Bigr] \\
	\displaystyle\int_{-\infty}^\infty\!\! d\Omega\, \Bigl[ a^\text{out}_\Omega \varphi^*_\Omega(x,t)
				\!+\! {{}\hat{a}^\text{out}_\Omega}^\dagger \varphi_{-\Omega}(-x,t) \Bigr]
	\end{cases} \\
\Psi(x,t) &= 
	\begin{cases}
	\displaystyle\int_{-\infty}^\infty\!\! d\bar\Omega\ \Bigl[ b^\text{in}_{\bar\Omega} \varphi_{\bar\Omega}(x,t)
			+ {{}\hat{b}^\text{in}_{\bar\Omega}}^\dagger \varphi^*_{-\bar\Omega}(-x,t) \Bigr] \\
	\displaystyle\int_{-\infty}^\infty\!\! d\bar\Omega\ \Bigl[ b^\text{out}_{\bar\Omega} \varphi^*_{\bar\Omega}(x,t)
			\!+\! {{}\hat{b}^\text{out}_{\bar\Omega}}^\dagger \varphi_{-\bar\Omega}(-x,t) \Bigr]
	\end{cases} \\
\psi(x,t) &= \int_{-\infty}^\infty dk\ \Bigl[ c_k e^{-i\omega(k) t+ikx} + c^\dagger_k e^{i\omega(k) t - ikx} \Bigr]
\end{align*}
The upper and lower expansions are to be taken at early and late times, respectively.
The vacua of the three sectors are defined in the usual way,
$a^\text{in}\ket{0_\Phi^\text{in}}=0$, $b^\text{in}\ket{0_\Psi^\text{in}}=0$ and
$a^\text{out}\ket{0_\Phi^\text{out}}=0$, $b^\text{out}\ket{0_\Psi^\text{out}}=0$
for early and late times, respectively, and $c\ket{0_\varphi}=0$ for the test field.

The transition amplitude for the $\Phi\to\Psi\psi$ process is now readily computed using
the standard $S$-matrix formalism in the coupling $\epsilon$ (upper in/out indices)
combined with the in-out formalism for the $\Phi$ and $\Psi$ fields (lower indices):
\begin{align}
&\abs{\mathcal{A}_{\Phi\rightarrow\Psi\psi}}^2 \nonumber\\
	&\quad= \abs{\bra{0^\text{out}}\bigl(b^\text{out}_{\bar\Omega} c_k\bigr)_\text{out}
			\bigl({a^\text{in}_\Omega}^\dagger\bigr)_\text{in}\ket{0^\text{in}}}^2 \label{eq:AmplitudeAnsatz}\\
	&\quad= \epsilon^2 \abs{ 2\pi \delta(\Omega-\bar\Omega-\omega(k))
			\int_{-\infty}^{\infty} dx\ e^{-ikx} \varphi_\Omega(x) \varphi_{\Bar\Omega}(x) }^2 \nonumber
\end{align}
We use the integral representation of the parabolic cylinder functions
\begin{equation}
E_\mu(x)=e^{i\delta} \frac{\sqrt{2}e^{\frac{\pi\mu}{4}}}{\Gamma\bigl(i\mu+\frac{1}{2}\bigr)}
	\int_0^\infty\! ds\, s^{i\mu-\frac{1}{2}} e^{\frac{1}{4}ix^2-\sqrt{-i}xs-\frac{1}{2}s^2}
\end{equation}
so that the integral over $x$ is an ordinary Gaussian integral which is readily computed:
\begin{widetext}
\begin{multline}\label{eq:xIntegral}
\abs{ \int_{-\infty}^{\infty} dx\ e^{-ikx} \varphi_\Omega(x) \varphi_{\Bar\Omega}(x) }^2 \\
= \abs{2\beta\bar\beta}^2 \frac{e^{\pi(\mu+\bar\mu)}}{\textstyle\abs{\Gamma\bigl(i\mu+\frac{1}{2}\bigr)}^2
			\abs{\Gamma\bigl(i\bar\mu+\frac{1}{2}\bigr)}^2}\ \frac{\pi}{8E^2}\,
	\Biggl| \int_0^\infty ds\ s^{-i\mu-\frac{1}{2}}\ e^{-i\frac{k-\Omega+\bar\Omega}{\sqrt{2E}}s}\ 
			\int_0^\infty dt\ t^{-i\bar\mu-\frac{1}{2}}\ e^{-i\frac{k+\Omega-\bar\Omega}{\sqrt{2E}}t}\ e^{ist}
			\Biggr|^2
\end{multline}
\end{widetext}
The energy-conserving delta function~\eqref{eq:AmplitudeAnsatz} restricts the momentum $k$ to
$k=\pm\omega=\pm(\Omega-\bar\Omega)$, thus one of the exponentials will always result to a factor of one.
The integrals over $s$ and $t$ are both integral representations of Gamma functions, so that in the ``plus'' case
\begin{align}
\text{\eqref{eq:xIntegral}}&=\abs{2\beta\bar\beta}^2 \frac{e^{3\pi\mu}}
		{\textstyle\abs{\Gamma\bigl(i\bar\mu+\frac{1}{2}\bigr)}^2}\ \frac{\pi}{8E^2}\ 
		\abs{\Gamma\bigl(i(\bar\mu-\mu)\bigr)}^2 \\
&= \frac{\pi}{2E^2(\bar\mu-\mu)}\ \frac{\abs{\bar\alpha}^2}{e^{2\pi(\bar\mu-\mu)}-1}
\end{align}
(in  the ``minus'' case, there is an $\alpha$ instead of $\bar\alpha$ in the numerator).

Thus the probability for the transition $\Psi\to\Psi\psi$ is proportional to
\begin{equation}
P_{\Phi\to\Psi\psi} \propto \frac{\epsilon^2}{M\omega_d}\,\frac{\abs{\bar\alpha}^2}{e^{2\pi(\bar\mu-\mu)}-1}.
\end{equation}
We recognize the well-known Bose-Einstein factor
\[\frac{1}{e^{2\pi(\bar\mu-\mu)}-1} = \frac{1}{e^{\frac{2\pi}{a}\omega_\text{d}}-1}\]
at temperature $T=\frac{a}{2\pi}$ with the detector's energy gap $\omega_\text{d}=a(\bar\mu-\mu)$ (see eq.~\eqref{eq:MassGap})
and the acceleration expressed through the external electric field, $a=E/M$.
This agrees with the result obtained from the the Unruh-DeWitt model~\cite{Unruh76,DeWitt79,Audretsch94a} except for the factor
$\abs{\bar\alpha}^2$ (or $\abs{\alpha}^2$, depending on the sign of the momentum $k$) in the numerator;
this factor increasing the transition probability for strong accelerations (compared to the mass)
is to be read as correction term for the case of a ``real'' detector, as opposed to an ``ideal'' detector
which is infinitely heavy.

The relation between the energy $\omega$ of the emitted particle and the the detector's energy gap $\omega_\text{d}$
is again obtained using the energy and momentum conservation relation,
although we have to adopt a picture of localized wave packets since the momenta
\begin{subequations}\begin{align}
K(x)&=\pm\sqrt{(\Omega+Ex)^2-M^2} \label{eq:LocalMomentum}\\ \bar K(x)&=\pm\sqrt{(\bar\Omega+Ex)^2-\bar M^2}
\end{align}\end{subequations}
of the $\Phi$ and $\Psi$ particles are only defined locally.
The energy of the created $\psi$ particle then computes to
\begin{equation}\label{eq:EmissionEnergy}
\omega = \frac{K(x)-(\Omega+Ex)}{M}\,\omega_\text{d},
\end{equation}
where $x$ is to be taken at the coordinate where the particle is created.
It can be seen immediately from \eqref{eq:LocalMomentum} that $\omega$ is positive if and only if
\[x\le-\frac{M+\Omega}{E},\]
which we recognize as the turning point of a detector antiparticle of energy $\Omega$:
The $\psi$ particle is actually created beyond the turning point of the detector,
in the realm of an antiparticle of the same energy.

The redshift factor in \eqref{eq:EmissionEnergy} ensures that the detector mesures frequencies with respect to its
proper time regardless of the reference frame we are working in. Unlike in the conventional Unruh-DeWitt model,
this is not assumed a priori. Instead, the energy redshift is a direct consequence of energy and momentum conservation.

\subsection{Test particle absorption\label{sec:Absorption}}
The transition probability for excitation of the detector due to absorption of a test particle is computed in the same way.
The result is
\begin{equation}
P_{\Phi\psi\to\Psi} \propto \frac{\epsilon^2}{M\omega_d}\,\biggl[1+\frac{\abs{\bar\alpha}^2}{e^{2\pi(\bar\mu-\mu)}-1}\biggr].
\end{equation}
In addition to the Bose-Einstein term we get another term independent from the acceleration $a$ or the external field $E$.
This term remains in the limit $a\to0$ and corresponds to the detection process without the presence of an external acceleration.
We do the analysis of the energy-momentum conservation in analogy to \eqref{eq:EmissionEnergy} to find
\begin{equation}\label{eq:AbsorptionEnergy}
\omega = \frac{(\Omega+Ex)-K(x)}{M}\,\omega_\text{d},
\end{equation}
so that $\omega$ is positive if the absorption happens in the realm of the detector particle, $x\ge\frac{M-\Omega}{E}$.
This result again agrees with the localized discussion of the Unruh-DeWitt detector by Audretsch and M\"uller~\cite{Audretsch94a}
(except for the factor of $\abs{\bar\alpha}^2$ in front of the Bose-Einstein factor).

\section{Summary}
In this paper, we have set up a relativistic Unruh detector accelerated by a constant external field
and found out that the detector becomes excited even in Minkowski vacuum. An observer looking just at the detector
(not at the test field) would describe this as ``detection'' of particles. The probability
for this ``detection'' follows a thermal Bose-Einstein distribution of temperature $T=\frac{a}{2\pi}$ in the
detection energy, $\omega_\text{d}=\frac{\bar M^2-M^2}{2M}$. The conventional interpretation is
that this effect comes from the thermal bath of Rindler particles the detector is immersed in due to the Unruh effect.
However, as we encountered, in this model the process involved is in fact spontaneous emission of test particles together with
spontaneous excitation of the detector. This is the picture from the point of view of an inertial observer
(such as ourselves, at least in good approximation)
who is unaware of the Rindler reference frame and the particle definition associated with it: 
He observes that the detector has switched to its excited state and, in the same process, has emitted a test particle.

The quantum field theoretical approach has two main benefits over the conventional Unruh-DeWitt ``particle in a box'' model.
The first one is that this approach works without the assumption that a non-inertial detector measures
frequencies with respect to its proper time~\cite{Sciama81}:
In the conventional model, the acceleration of the detector is encoded only
in the relation between proper time $\tau$ and Minkowski time $t=t(\tau)$, which is the crucial part of the interaction term
for obtaining the thermal distribution in the Green's functions and Bogolyubov coefficients.
In the model discussed here, the interaction is given in ordinary Minkowski coordinates, while the
acceleration is implemented in the action of the detector itself by incorporating the \emph{source} of the acceleration,
the background field (which is entirely missing in the Unruh-DeWitt model); the relation between the energy $\omega_d$
as measured by the detector and the energy $\omega$ as measured by an inertial observer then simply follows from
energy and momentum conservation
(see equations \eqref{eq:InertialEnergy}, \eqref{eq:EmissionEnergy} and \eqref{eq:AbsorptionEnergy}).

The other advantage of this approach is that it does not rely on the notion of Rindler particles.
This solves the problem with inequivalent particle definitions: We can work in Minkowski vacuum,
which is the only vacuum state compatible with general relativity and therefore has to be seen as the ``true'' physical vacuum.
DeWitt~\cite{DeWitt79} views the reaction of the accelerated detector to Minkowski vacuum as an ``operational definition''
for Rindler particles. Following our point of view, we read the spontaneous excitation of the detector as the actual
physical process, while we consider the Rindler vacuum only as an auxiliary device to describe the reaction of
an accelerated detector without having to explicitly specify the source of the acceleration.

\bibliography{references}

\end{document}